\documentclass[12]{article} 
\usepackage{natbib}
\usepackage{graphicx}
\usepackage{amsmath}
\usepackage[top=1in,bottom=1in,left=0.75in, right=0.75in]{geometry}
\usepackage{authblk}
\usepackage{indentfirst}
\usepackage[titletoc,title]{appendix}
\usepackage{tikz}
\usepackage{breqn}
\usepackage{mathtools}
\usepackage{subfig}
\usepackage{tabularx}
\usepackage{float}
\usepackage{setspace}
\usepackage{caption}
\usepackage[font=footnotesize]{caption}
\usetikzlibrary{shapes.geometric, arrows}
\date{}

\usetikzlibrary{shapes.geometric, arrows}
\date{}
\newtheorem{myDef}{Definition} 
\newtheorem{myDef1}{Definition}
\newtheorem{theo}{Theorem} 
\newtheorem{pr}{Proof} 
\newtheorem{rem}{Remark} 

\title{\bf Stability analysis of stochastic second-order macroscopic continuum models and numerical simulations}

\author[a]{\textsc{\small Marouane Bouadi}}
\author[a]{\textsc{\small Bin Jia}\thanks{Corresponding authors: bjia@bjtu.edu.cn (Bin Jia), jiangrui@bjtu.edu.cn (Rui Jiang)}}
\author[a]{\textsc{\small Rui Jiang*}}
\author[a]{\textsc{\small Xingang Li}}
\author[a]{\textsc{\small Zi-You Gao}}

{\affil[ \small  a]{\small Key Laboratory of Transport Industry of Big Data Application Technologies for Comprehensive Transport, Ministry of Transport, Beijing Jiaotong University, Beijing 100044, China}

\begin{document}

\maketitle

\abstract{
Second-order macroscopic continuum models have been constantly improving for decades to reproduce the empirical observations. Recently, a series of experimental studies have suggested that the stochastic factors contribute significantly to destabilizing traffic flow. Nevertheless, the traffic flow stability of the stochastic second-order macroscopic continuum model hasn't received the attention it deserves in past studies. More importantly, we have found that the destabilizing aspect of stochasticity is still not correctly validated in the existing theoretical stability analysis. In this paper, we analytically study the impact of stochasticity on traffic flow stability for a general stochastic second-order macroscopic model by using the direct Lyapunov method. Numerical simulations have been carried out for different typical stochastic second-order macroscopic models. Our analytical stability analysis has been validated, and our methodology has been proved more efficient. Our study has theoretically revealed that the presence of stochasticity has a destabilizing effect in stochastic macroscopic models.


}\hspace{10pt}


\section{Introduction}

In vehicular traffic science and engineering, scholars have developed many theories to explain the complex traffic flow dynamics. Many fascinating non-linear traffic phenomena have been reproduced and studied, such as, the dynamics of phantom jam emergence \citep{gazis}, traffic oscillations \citep{mauch}, the dynamics of traffic humps \citep{light}, metastability \citep{kerner}, traffic stability under small perturbations \citep{bando}, see more details in \cite{trbb} and \cite{sha}. From the practical perspective, recent studies aimed to propose efficient control strategies, e.g. trajectory smoothing methods \citep{yao,chen,han}, to reduce or eliminate the emergence of traffic oscillations; hence saving fuel consumption and reducing time delay \citep{lix}. 

To understand the generation mechanism of traffic instability, extensive research studies have been carried out on both microscopic and macroscopic traffic models. The stability analysis methods include the frequency domain analysis for microscopic models \citep{her,mas,plo,mont}, the wave expansion technique for macroscopic models \citep{wex,jing}, and the indirect Lyapunov method based on eigenvalue analysis for both microscopic and macroscopic models \citep{bando,ward,wil,trbb,zh}. The derived stability conditions mostly depend on the speed adaptation time, traffic density, and drivers' sensitivity to the gap and velocity difference. However, only the deterministic traffic dynamics have been considered. 


In literature, various stochastic traffic models have been proposed. To our knowledge, the first developed stochastic model is the Nagel and Schreckenberg (NaSch) cellular automaton model \citep{ns}. It is found that the mechanism of vehicles' spontaneous deceleration can reproduce the appearance of phantom jams. More recently, \cite{boel} and \cite{sumale} have extended the cell transmission model (CTM) proposed by \cite{daganzo} to take into account uncertainties in demand and supply (SCTM) on freeways. \cite{jabari,jabari2} have developed a CTM model with uncertainty in drivers' gap choice, which has the advantage of avoiding negative paths. The fundamental diagram subject to the presence of noise has been investigated as well. For instance, \cite{jiali} have studied an LWR model with uncertainty in the free flow speed, \cite{ngod} has studied a multiclass LWR model with uncertainty in traffic capacity and reproduced the wide scattering in the fundamental diagram, \cite{qian} have studied a mesoscopic model with uncertainty on transition rates and vehicles states, \cite{elv} have studied an LWR model with stochastic traffic parameters including the maximum velocity, the maximum density, and transitions between different traffic regimes in the velocity-density plane. \cite{wang} used the Kalman filter to estimate traffic state by using a stochastic second-order traffic model. Details about the traffic flow dynamics with stochasticity can be found in the book of \cite{chen2}. Nevertheless, the impact of drivers' uncertainties or stochasticity on stabilizing or destabilizing traffic systems has not yet gotten the attention it deserves.  

Recently, \cite{jiang2,jiang1} carried out a series of experiments suggesting that speed plays a critical role in traffic stability. In this context, \cite{jiang1} have shown that a critical speed exists below which traffic is unstable. Moreover, \cite{jiang1} have suggested that the traffic instability mechanism is probabilistic and results from a competition between stochastic disturbances and the speed adaptation effect. 

Stability analysis of stochastic traffic models is still scarce in the literature and represents an effervescent research direction. In this respect, recent stability analysis of microscopic models has been performed \citep{trb2,trb3,laval,ngod2,laval2,wang2,yuan, bouadi}. It was found that stochasticity can indeed destabilize traffic systems \citep{trb2,laval,ngod2,laval2}, and leads to the empirically observed concave growth pattern of the cars' speed standard deviation \citep{wang2,yuan,tian4}. Regarding macroscopic models, \cite{shiteng} have been the first to investigate the stability properties of a stochastic LWR model and a stochastic speed gradient model. To our knowledge, \cite{shiteng} were the first to validate the concave growth pattern from a macroscopic level. A quantitative difference has been observed for model stability as well. However, an analytical treatment to deeply understand the effect of stochasticity on traffic stability has not been yet performed.

More recently, \cite{ngodu} has analytically investigated the stability of a general class of stochastic macroscopic models where a stability condition has been derived. Nevertheless, there are two serious deficiencies in the study.
(i) From the methodological perspective, Eigenvalue method has been used in \cite{ngodu}. In this context, \cite{ngodu} argued that a stochastic differential equation of the type $dX=AXdt + RXdW$ (see equation (18) in \cite{ngodu}) is almost surely linearly stochastically stable if $\Re{(A- 0.5R^{2})}\leq 0$, where $X$ is the vector including the perturbed density and velocity for macroscopic traffic models, i.e. $X=[\tilde{\rho} \ \ \ \tilde{v}]^{T}$. However, the above stability condition has been proved valid only when the matrices A and R commute \citep{maob}, which is usually not the case in stochastic macroscopic models. (ii) As a result of (i), the stochasticity can sometimes stabilize traffic flow which is in disagreement with empirical findings and also is against Proposition 1 in \cite{ngodu} as we will see in this study. This work attempts to address the previously mentioned deficiencies.


In this paper, we carry out an analytical stability analysis of a general class of stochastic second-order macroscopic models by using the generalized Lyapunov equation. Our analytical stability analysis will enable us to study the stability properties of different traffic models in the presence of stochasticity. We will see that the stochastic macroscopic traffic models have interesting properties that have not been investigated in previous studies. Moreover, we will also see that our methodology resolves the deficiency in \cite{ngodu} in deriving stability conditions of stochastic second-order macroscopic models in general. To validate our theoretical analysis, numerical simulations will be carried out for four typical macroscopic models, namely, the Aw-Rascle model \citep{aw}, the speed gradient model proposed \cite{jiang5}, the Zhang model \citep{zhan} and the GKT model proposed by \citep{trb}.  

Our paper will be organized as follows; in the following section, we will briefly review the general expression of stochastic second-order macroscopic models. Next, we perform the corresponding stability analysis. Afterward, we carry out numerical simulations for the four typical macroscopic models. The last section will be devoted to a conclusion.


\section{The stochastic second-order macroscopic model}

Empirical observations are always revealing that traffic flow is characterized by a certain degree of uncertainty in time and space due to the stochastic nature of human drivers. After conducting a series of experiments, \cite{jiang1} have demonstrated that traffic instability has a probabilistic nature. Moreover, \cite{jiang1} found that the presence of stochastic factors has a destabilizing effect, and the generation mechanism of traffic instability results from a competition between the speed adaptation effect and the stochastic factors. Consequently, the existing second-order macroscopic models need to be improved by explicitly considering the presence of stochastic factors.  

\subsection{The stochastic model}


The general stochastic second-order stochastic model reads \citep{trbb, ngodu}:


\begin{equation}
\frac{\partial\rho}{\partial t}+ \frac{\partial (\rho v)}{\partial x}=0,
\end{equation}

\begin{equation}
\frac{\partial v}{\partial t} + v \frac{\partial v}{\partial x}= f_{1}(\rho,v,\frac{\partial \rho}{\partial x},\frac{\partial v}{\partial x},\rho_{a},v_{a})+f_{2}(\rho,v),
\end{equation}

\noindent Equations (1) and (2) are the density and the velocity equation, respectively. $f_{1}$ is the deterministic term of equation (2), and $f_{2}$ is the stochastic term which has the form $f_{2}(\rho,v) dt=h(\rho,v) dW$, where $h$ is a regular function and $dW$ denotes the Brownian motion. In the velocity equation, the deterministic term on the right-hand side encompasses the interaction term and a relaxation term reflecting the vehicles' tendency to reach a desired speed. The terms $\rho_{a}=\rho(x+d,t)$ and $v_{a}=v(x+d,t)$ are respectively the density and the velocity evaluated at a distance $x+d$. Those terms account for the interaction distance in the non-local Gas Kinetic Traffic (GKT) model \citep{trb}. Due to the uncertain nature of human drivers and the estimation errors, the velocity equation (2) has been generalized by considering the presence of stochastic factors.


\subsection{Existence and uniqueness of the solution}

Since we are dealing with a stochastic partial differential equation, it would be interesting to check the condition of existence and uniqueness of the solution. To this aim, we apply the definition of existence and uniqueness to the stochastic continuum macroscopic model in equations (1) and (2), See Appendix A.  It can be easily shown that, for regular functions $f_{1}$ and $f_{2}$, the solution exists and is unique, if (i) $\lVert f_{2} \rVert \leq \sigma \lVert x \rVert $, (ii) $\sigma \leq C$, where $\lVert . \rVert$ is the Euclidian norm, $x=(\rho,v)$, $\sigma$ is a constant and $C$ denotes the upper bound of the Euclidian Norm of the deterministic term in equations (1) and (2).


\subsection{Linearization}

To carry out the stability analysis, we consider a circular road where the density of vehicles is conserved. In this study, we study traffic stability in the case of a small perturbation around the equilibrium. We linearize the above-defined density and velocity equations around the equilibrium density $\rho_{e}$ and velocity $v_{e}$; hence, equations (1) and (2) become:
 
\begin{equation}
\frac{\partial  \tilde{\rho}}{\partial t} + \rho_{e} \frac{\partial  \tilde{v}}{\partial x} + v_{e} \frac{\partial  \tilde{\rho}}{\partial x} =0, 
\end{equation}

\begin{equation}
\frac{\partial  \tilde{v}}{\partial t} + v_{e} \frac{\partial  \tilde{v}}{\partial x} =\tilde{v} \frac{\partial f_{1}}{\partial v}+\tilde{\rho} \frac{\partial f_{1}}{\partial \rho} + \frac{\partial  \tilde{v}}{\partial x} \frac{\partial f_{1}}{\partial d_{v}} + \frac{\partial \tilde{\rho}}{\partial x} \frac{\partial f_{1}}{\partial d_{r}} + \tilde{v}_{a} \frac{\partial f_{1}}{\partial v_{a}}+ \tilde{\rho}_{a} \frac{\partial f_{1}}{\partial \rho_{a}}+(\mu \tilde{\rho}+\eta \tilde{v}) \xi(t), 
\end{equation}

\noindent where

\begin{equation}
\rho = \rho_{e} + \tilde{\rho},
\end{equation}

\begin{equation}
v = v_{e} + \tilde{v},
\end{equation}

\noindent and $d_{v}=\frac{\partial v}{\partial x}$, $d_{r}=\frac{\partial \rho}{\partial x}$. For conciseness, the following notations will also be adopted $f_{1v}=\frac{\partial f_{1}}{\partial v}$, $f_{1\rho}=\frac{\partial f_{1}}{\partial \rho}$,  $f_{1vx}=\frac{\partial f_{1}}{\partial d_{v}}$, $f_{1\rho x}=\frac{\partial f_{1}}{\partial d_{r}}$, $f_{1va}=\frac{\partial f_{1}}{\partial v_{a}}$, $f_{1\rho a}=\frac{\partial f_{1}}{\partial \rho_{a}}$, $\mu=\frac{\partial f_{2}}{\partial \rho}$ and $\eta=\frac{\partial f_{2}}{\partial v}$.

 Supposing that the perturbations are given by wave equations, we consider the following Ansatz  $\tilde{\rho} = \hat{\rho}(t,\omega) e^{-ikx}$ and $\tilde{v} = \hat{v}(t,\omega) e^{-ikx}$, where $\omega$ is a random event. Note that in this case, we have $\tilde{\rho}_{a}=\tilde{\rho} e^{-ikd}$ and $\tilde{v}_{a}=\tilde{v}e^{-ikd}$. 

To simplify the analysis, we perform a first order Taylor approximation (long wave length approximation) to the previous interaction terms, i.e. $\tilde{\rho}_{a}=\tilde{\rho} (1-ikd)$ and $\tilde{v}_{a}=\hat{v}(1-ikd)$.  Hence, we obtain the following stochastic differential equation:

\begin{equation}
dx=Axdt+RxdW,
\end{equation}

\noindent where

\begin{equation}
x = [\tilde{\rho} \ \ \ \tilde{v}]^{T},
\end{equation}

\noindent and 

\begin{equation}
A= {\begin{bmatrix}
    ikv_{e}       & ik\rho_{e}  \\
    f_{1\rho}+f_{1\rho a}-ikf_{1\rho x}-ikdf_{1\rho a}  & f_{1v}+f_{1va}+ikv_{e}-ikf_{1vx}-ikdf_{1va} \\
\end{bmatrix}}, 
\end{equation}

\begin{equation}
R= {\begin{bmatrix}
    0       & 0  \\
  \mu  & \eta\\
\end{bmatrix}}.
\end{equation}

Equation (7) is an autonomous stochastic differential equation for which we will derive a stability condition in the following section. To derive a closed stability condition, we suppose that $f_{2}$ depends only on velocity \citep{ngod,xu}; hence $\mu=0$. For a function $f_{2}$ that depends on both the density and the velocity, a numerical analysis will be conducted to search the Lyapunov function and prove stability.

\section{Traffic flow stability analysis}

In this section, we extract a stability condition of equation (7) in the mean square sense. For this purpose, we first recall some important results of the stability of stochastic differential equations. 

\subsection{Stability of stochastic differential equations}

In this work, we are interested in the stability of macroscopic traffic flow models. They are two widely used methods to study the stability of physical systems for linear and deterministic differential equations of the form $x(t)= Ax$: the direct Lyapunov method and the eigenvalue method (often called the indirect Lyapunov method). In the eigenvalue method, the stability problem depends on the eigenvalues of matrix A. Subsequently, the stability is guaranteed when the real parts of the eigenvalues are negative. The eigenvalue method has been extensively used for studying the stability of deterministic traffic systems, see for example \cite{trbb}.  
However, except in some special cases, stochastic differential equations of the form $dx= Ax+RdW$, e.g. equation (7), do not have explicit solutions as deterministic systems. Hence, there is no general method to study the stability problem by the eigenvalue method as deterministic systems. In this case, the direct Lyapunov method, which does not require an explicit solution to the problem, represents an alternative to overcome the issue.  

In \cite{ngodu}, the eigenvalue method has been used to derive a stability condition of the stochastic continuum second order traffic model, See definition 2 in \cite{ngodu}. However, we argue that the methodology followed in \cite{ngodu} is not suited for studying the stability of stochastic differential equations. According to definition 2, the linear stochastic differential equation is said to be almost surely linearly stochastically stable if:

\begin{equation}
\Re{(A^{*})}=  \Re{(A- 0.5R^{2})}\leq 0
\end{equation}

In the absence of a proof, equation (11) supposes that equation (7) does have an explicit solution, which is not the case. In fact, according to \cite{maob}, equation (11) holds only for a special case where the matrices A and R commute, See example 3.8 chapter 4 in \cite{maob}. Hence, the stability of stochastic macroscopic differential equations needs to be revisited. 

In our context, the explicit solution of the stochastic differential equation is difficult to extract, especially in the presence of complex-valued matrices, see equation (7). Hence, we believe that the direct Lyapunov method represents the best alternative. In the following, we will compare the result of our analysis with the result obtained in \cite{ngodu} to demonstrate the superiority of our approach. 

\subsection{Mean-square stability of stochastic differential equations}

The mean-square stability plays an important role in control theory. In the context of stochastic differential equations, the second moment of the solution $x(t)$ in equation (7) will tend to zero. The definition of the asymptotic mean-square stability is given by the following.

\begin{myDef}
     A stochastic system of the form (7) is said to be asymptotically mean square stable if: 
		
	\begin{equation}
\lim_{t\to \infty} E{||x(t)||}^{2}=0  \nonumber,
\end{equation}

\noindent For any initial state $x(0) \in \Re^n$. 

\end{myDef}

The Lyapunov formalism represents a powerful tool to prove the asymptotic stability of a given stochastic differential equation without knowing the corresponding explicit solution. In the following, we will study the mean-square stability of equation (7) which has a complex valued matrix A. 

\begin{theo}

Consider a quadratic Lyapunov function $V(x)=x^T P x$, the autonomous stochastic differential equation (7) is mean-square stable if we define a positive function $V(x)$ such that $LV(x)\leq 0$ \citep{zhang3}.
\end{theo}

In this case, the matrix $P$ is positive definite and $LV(x)$ is given by: 


\begin{equation} 
LV(x)=x^{T} (PA+A^{T} P+R^T PR)x.
\end{equation}


\noindent Equation (12) is related to the differential $dV(x,t)$ by the following relation (Ito formula):

\begin{equation} 
dV(x,t)=LV(x)dt+V_{x}(x)h(x)dW, 
\end{equation}

\noindent where $V_{x}(x)$ denotes derivative of $V(x)$ with respect to $x$. $h(x)$ is the stochastic component of equation (7). The above theorem only deals with real stochastic system while equation (7) is a complex valued stochastic system. In the following, we will present a method that yields the equivalent real system.

\subsection{Stability analysis of the general stochastic macroscopic model}

In this subsection, we exploit the previously discussed definition of the asymptotic mean-square stability to extract a closed form stability condition of equation (7) where the stochastic term, i.e. $f_{2}$, depends only on the velocity, i.e. $\eta\neq 0$ and $\mu=0$. 

For a stochastic term that depends on both the density and the velocity, i.e. $\eta\neq 0$ and $\mu\neq 0$, the extraction of a closed form stability condition becomes cumbersome. However, the stability issue can be solved numerically. 

\subsubsection{Velocity dependent stochastic term}

For a velocity dependent stochastic term, an approximated stability condition of the general stochastic second-order macroscopic model (equations (1) and (2)) is given by the following theorem:

\begin{theo}
 
The linearized stochastic differential equation given by equation (7) is mean-square stable if: 


\begin{equation} 
2 \rho_{e}-\frac{(\eta^{2}+2f_{1v}+2f_{1va})(f_{1\rho}f_{1vx}-f_{1\rho x}f_{1v}+f_{1\rho a}f_{1vx}-f_{1\rho x}f_{1va}+d(f_{1\rho}f_{1va}-f_{1\rho a}f_{1v}))}{(f_{1\rho}+f_{1\rho a})^2} \leq 0.
\end{equation}

\end{theo}

\begin{pr}

The linearized form of equations (1) and (2) is given by the linear stochastic differential equation (7) which has a complex valued matrix $A$. To apply the Lyapunov method, we first opt by separating the real and imaginary part of the system (7), that is, $x_{s}=x_{r}+ix_{i}$, where the vectors $x_{r}$ and $x_{i}$ are respectively the real part and the imaginary part of the solution x in the system (7). Hence, the system (7) is equivalent to the following system \citep{zhang1}: 

\begin{equation}
dx_{s}=A_{s}x_{s}dt+R_{s}x_{s}dW,
\end{equation}

\noindent where 

\begin{equation}
x_{s} = [x_{r}\ \ \ x_{i}]^{T},
\end{equation}

\noindent and

\begin{equation}
A_{s}= {\begin{bmatrix}
    0      & 0 & -k v_{e}& -k\rho_{e}\\
   f_{1\rho}+f_{1\rho a}   &  f_{1v}+f_{1va} & k(f_{1\rho x} +d f_{1\rho a}) & k(f_{1vx}+d f_{1va}-v_{e}) \\
	 k v_{e}   &  k\rho_{e} & 0 & 0\\
	 -k(f_{1\rho x} + d f_{1\rho a})   &  k(v_{e}-f_{1vx}-d f_{1va})  & f_{1\rho}+f_{1\rho a} & f_{1v}+f_{1va}\\
\end{bmatrix}}, 
\end{equation}

\begin{equation}
R_{s}= {\begin{bmatrix}
    0      & 0 & 0 & 0 \\
  0   &  \eta  & 0 & 0 \\
	 0   &  0 & 0 & 0\\
	 0    & 0 & 0 & \eta \\
\end{bmatrix}}, 
\end{equation}

\noindent Then, we consider the following quadratic Lyapunov function $V(x_{s})$ which must be a positive function:

\begin{equation} 
V(x_{s})=x_{s}^{T}Px_{s},
\end{equation}

\noindent Consequently, for the stochastic system (7), the operator $LV(x)$ in equation (12) will be given by: 

\begin{equation} 
LV(x_{s})=x_{s}^{T} (PA_{s}+A_{s}^{T} P+R_{s}^T PR_{s})x_{s}.
\end{equation}

\noindent The stochastic system (7) is stable if the operator $LV(x_{s})$ has negative eigenvalues:

\begin{equation} 
PA_{s}+A_{s}^{T} P+R_{s}^T PR_{s} \leq 0. 
\end{equation}

\noindent Next, we should define a positive definite matrix $P$ such that equation (21) is negative definite and the resulting stability condition is independent of the wave number $k$. To perform this task, we consider a matrix $P$ with unknown coefficients in the following equation: 


\begin{equation} 
T= PA_{s}+A_{s}^{T} P. 
\end{equation}

\noindent then, after simple eliminations and replacements to have a diagonal matrix $T$, we get the following matrix P:

\begin{equation}
P= {\begin{bmatrix}
    P_{1}      & P_{2} \\
  P_{2}^{T}   &  P_{1}  \\
	 
\end{bmatrix}}, 
\end{equation}

\noindent where the symmetric matrix $P_{1}$ reads:


\begin{equation}
P_{1}= {\begin{bmatrix}
    -\frac{(f_{1v}+f_{1va})(f_{1\rho}^2+f_{1\rho a}^2+d^2 f_{1\rho a}^{2}k^{2}+k^{2}f_{1\rho x}^{2}+2d f_{1\rho a}f_{1\rho x}k^{2})}{\rho_{e}k^2(f_{1\rho a}+f_{1\rho})^{2}}      & -\frac{f_{1\rho x}+d f_{1\rho a}}{f_{1\rho}+f_{1\rho a}} \\
  -\frac{f_{1\rho x}+d f_{1\rho a}}{f_{1\rho}+f_{1\rho a}}  &   -\frac{(f_{1\rho}f_{1vx}-f_{1\rho x}f_{1v}+f_{1\rho a}f_{1vx}-f_{1\rho x}f_{1va}+d(f_{1\rho}f_{1va}-f_{1\rho a}f_{1v}))}{(f_{1\rho}+f_{1\rho a})^2}  \\
	 
\end{bmatrix}}, 
\end{equation}

\noindent and the skew-symmetric matrix $P_{2}$ is given by the following:
\begin{equation}
P_{2}= {\begin{bmatrix}
    0      & -\frac{1}{k} \\
  \frac{1}{k}   &  0  \\
	 
\end{bmatrix}}, 
\end{equation}

\noindent After replacing in equation (20), we get the following expression: 

\begin{equation}
LV(x)= (2 \rho_{e}-\frac{(\eta^{2}+2f_{1v}+2f_{1va})(f_{1\rho}f_{1vx}-f_{1\rho x}f_{1v}+f_{1\rho a}f_{1vx}-f_{1\rho x}f_{1va}+d(f_{1\rho}f_{1va}-f_{1\rho a}f_{1v}))}{(f_{1\rho}+f_{1\rho a})^2} )(\tilde{v_{r}}^{2}+ \tilde{v_{i}}^{2}),
\end{equation}

\noindent where $\tilde{v}_{r}$ and $\tilde{v}_{i}$ are respectively the real and imaginary components of $\tilde{v}$ in equations (8) and (15).


\noindent The matrix P is positive definite if the following condition is met:

\begin{equation} 
-4k^{2}\rho_{e} q_{1} q_{2} >0,
\end{equation}

\begin{equation} 
q_{1}=d^{2}f_{1\rho a}^{2}k^{2} +2d f_{1\rho a} f_{1\rho x}k^{2}+f_{1\rho}^{2}+f_{1\rho a}^{2}+2f_{1\rho}f_{1\rho a}+f_{1\rho x}^{2}k^{2},
\end{equation}

\begin{equation} 
q_{2}=\rho_{e}-\frac{(f_{1v}+f_{1va})(f_{1\rho}f_{1vx}-f_{1\rho x}f_{1v}+f_{1\rho a}f_{1vx}-f_{1\rho x}f_{1va}+d(f_{1\rho}f_{1va}-f_{1\rho a}f_{1v}))}{(f_{1\rho}+f_{1\rho a})^2},
\end{equation}

\noindent The negativity of equation (14) implies the positivity of equation (27) with $\eta=0$. 

\noindent On the other hand, the negativity of equation (26) implies the stability condition in equation (14). 

\end{pr}

\noindent In the following, we draw some important remarks from the previously established stability condition. 

\begin{rem}

In the deterministic case $\eta=0$, we recover the stability condition of the deterministic traffic models \citep{trbb}:  

\begin{equation} 
2 \rho_{e}-\frac{(2f_{1v}+2f_{1va})(f_{1\rho}f_{1vx}-f_{1\rho x}f_{1v}+f_{1\rho a}f_{1vx}-f_{1\rho x}f_{1va}+d(f_{1\rho}f_{1va}-f_{1\rho a}f_{1v}))}{(f_{1\rho}+f_{1\rho a})^2}\leq 0.
\end{equation}

\end{rem}

\begin{rem}
Note that in real traffic, we have the following conditions $f_{1vx}>0$, $f_{1\rho x}<0$, $f_{1v}<0$, $f_{1\rho}<0$, $f_{1va}>0$ and $f_{1\rho a}<0$. Hence, the term multiplied with the noise's strength $\eta$ in equation (14) is always positive. Consequently, the stability condition (14) suggests that the stochastic nature of human of drivers tends to destabilize the traffic flow system. 
\end{rem}

\begin{rem}
For deterministic macroscopic traffic models, the stability condition in equation (30) has been obtained by eigenvalue analysis (indirect Lyapunov method). The present method is simpler and represents a fast way to obtain stability conditions of macroscopic models.    
\end{rem}

\subsubsection{Density and velocity dependent stochastic term}

The above stability analysis can be extended to a more general function $f_{2}(\rho,v)$ where $\eta\neq0$ and $\mu\neq0$. However, by following the same steps in the proof, one might get a large matrix P for which positivity would be difficult to demonstrate analytically. Hence, the stability condition corresponding to a density and velocity dependent stochastic term does not have a simple closed form. In this case, numerical analysis can be performed to search for the Lyapunov function.  

To this aim, we consider the matrix inequality equation (21) in which the matrix P is the unknown variable under the constraint that P is positive defined. In control theory, such optimization problems in which unknown variables are matrices are called Linear Matrix Inequalities (LMI) \citep{tob}. Our problem will hence take the form of the following convex optimization problem:    

\begin{equation}
{\begin{bmatrix}
    A_{s}^{T} P + PA_{s} + R^{T} P R      & 0 \\
  0   &  -P  \\
\end{bmatrix}}<0, 
\end{equation}

\noindent Equation (31) will be used to search for positive defined matrices P in the following numerical analysis.


\section{Numerical analysis of second order macroscopic traffic models}

In this section, we carry out numerical simulations of four typical stochastic second-order macroscopic traffic models, namely the stochastic Aw-Rascle model \citep{aw}, the stochastic speed gradient model \citep{jiang5}, the stochastic Zhang model \citep{zhan}, and the stochastic Gas Kinetic Traffic (GKT) model \citep{trb}. The previously mentioned traffic models can successfully avoid the wrong travel problem \citep{daganzo2, aw}. The first three models can additionally avoid the characteristic speed problem. Note that the first three models all have a speed gradient term and differ only in the propagation speed of small disturbances.  

To numerically study the traffic flow evolution, we will adopt the first-order upwind numerical integration scheme for the stochastic Aw-Rascle model, the stochastic speed gradient model, and the stochastic Zhang model (See an example in Appendix B). For these models, the following fundamental diagram will be adopted \citep{lee}: 

\begin{equation}
v_{e}(\rho)=\frac{v_{max}(1-\frac{\rho}{\rho_{max}})}{1+E(\frac{\rho}{\rho_{max}})^{4}},
\end{equation}

\noindent where E=100, $\rho_{max}=0.15 \ \mathrm{veh/m}$ and $v_{max}=30 \ \mathrm{m/s}$.

For the stochastic GKT model \citep{trb}, we will use the first-order upwind numerical integration adopted by \cite{ngodu}.

According to the definition of string instability, traffic is considered unstable if the amplitude of fluctuations grows with time along the lattice \citep{trbb}. Hence, to give an approximation of the numerical boundary, we ensure that: (i) the overall velocity's standard deviation in the second half of the simulation time exceeds the overall velocity's standard deviation in the first half \citep{shiteng} for multiple initial seeds, (ii) Appearance of stop and go waves in the space-time diagram.

Regarding the stochastic term, authors have proposed different expressions of the function $f_{2}$ in equation (2) \citep{xu,ngod2,ngodu}. In this work, we will study two different stochastic terms. The first one is the velocity dependent stochastic term, where the standard deviation increases when the velocity increases \citep{ngod2}:


\begin{equation}
f_{2}(v)= \sigma \sqrt{v} \xi(t).
\end{equation}

The second one is the density and velocity dependent stochastic term, which is given by \citep{ngodu}: 

\begin{equation}
f_{2}(\rho,v) = \sigma_{0} \frac{\rho}{\rho_{max}} (v_{0}-v) \xi(t),
\end{equation}

\noindent where $\sigma$ is the noise's strength or dissipation term, $\xi(t)$ is the white noise which is the derivative of the Wiener process $\xi(t)=\frac{dW(t)}{dt}$. For convenience, we will consider  $\sigma = \frac{\sigma_{0}}{\rho_{max}}$ when studying the second stochastic term. The following methodology can be used for any expression of the stochastic term $f_{2}$.

To carry out numerical simulations, we consider a road with periodic boundary conditions. The default value of the number of cells is 1000 cells where each cell has a width $\Delta x$=10 m (a road of length 10 km). The default value of the time step is $\Delta t$=0.05 s. 

\subsection{The stochastic Aw-Rascle model}

The velocity equation of the stochastic Aw-Rascle model \citep{aw} has the following expression:  

\begin{equation}
\frac{\partial v}{\partial t} + v \frac{\partial v}{\partial x} = \frac{v_{e}({\rho})-v}{\tau} + \rho P'(\rho) \frac{\partial v}{\partial x} + f_{2},
\end{equation}

\noindent where $P'(\rho)$ is the derivative of the traffic pressure, $\tau$ is the relaxation time and $v_{e}$ is the equilibrium velocity. According to \cite{aw}, the traffic pressure is assumed to have the following expression $P(\rho)\propto \rho^{\gamma}$, where $\gamma>0$. 




\underline{Stability condition}

To obtain the stability condition of the stochastic Aw-Rascle-Zhang supposing a velocity dependent stochastic term, i.e. equation (33), the following replacements can be made in equation (14): $f_{1vx}=\rho_{e}P'(\rho_{e})$, $f_{1\rho x}=0$, $f_{1\rho a}=0$, $f_{1va}=0$ ,$f_{1v}=-\frac{1}{\tau}$ and $f_{1\rho}=\frac{v'_{e}}{\tau}$. Thus, we get the corresponding stability condition: 
   
\begin{equation}
(2-\tau \eta^2)P'(\rho_{e})+2 v'_{e} \geq 0. 
\end{equation}

\noindent where $\eta= \frac{\sigma}{2\sqrt{v_{e}}}$. 
In the deterministic case ($\eta=0$), the above stability condition becomes $P'(\rho_{e})+v'_{e} \geq 0 $. 

For the density and velocity dependent stochastic term, i.e. equation (34), we follow the same methodology explained in the previous section by solving the LMI in equation (31) for a large parameter points and report the results in the $(\sigma,\rho_{e})$ plane. 



\underline{Numerical simulation}



\begin{figure}[H]
\centering
\subfloat[]{\includegraphics[width=.7\textwidth]{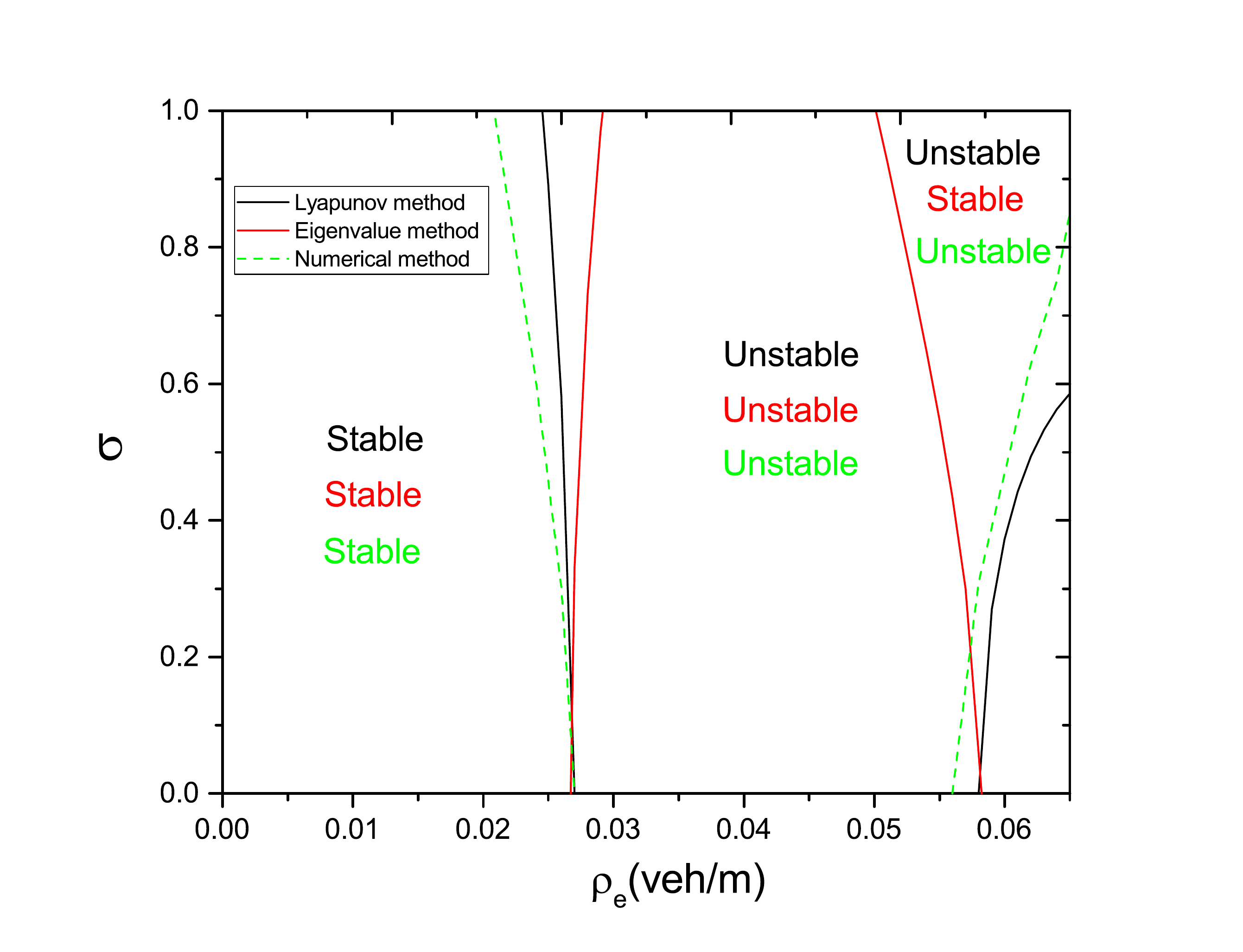}}\\
\subfloat[]{\includegraphics[width=.7\textwidth]{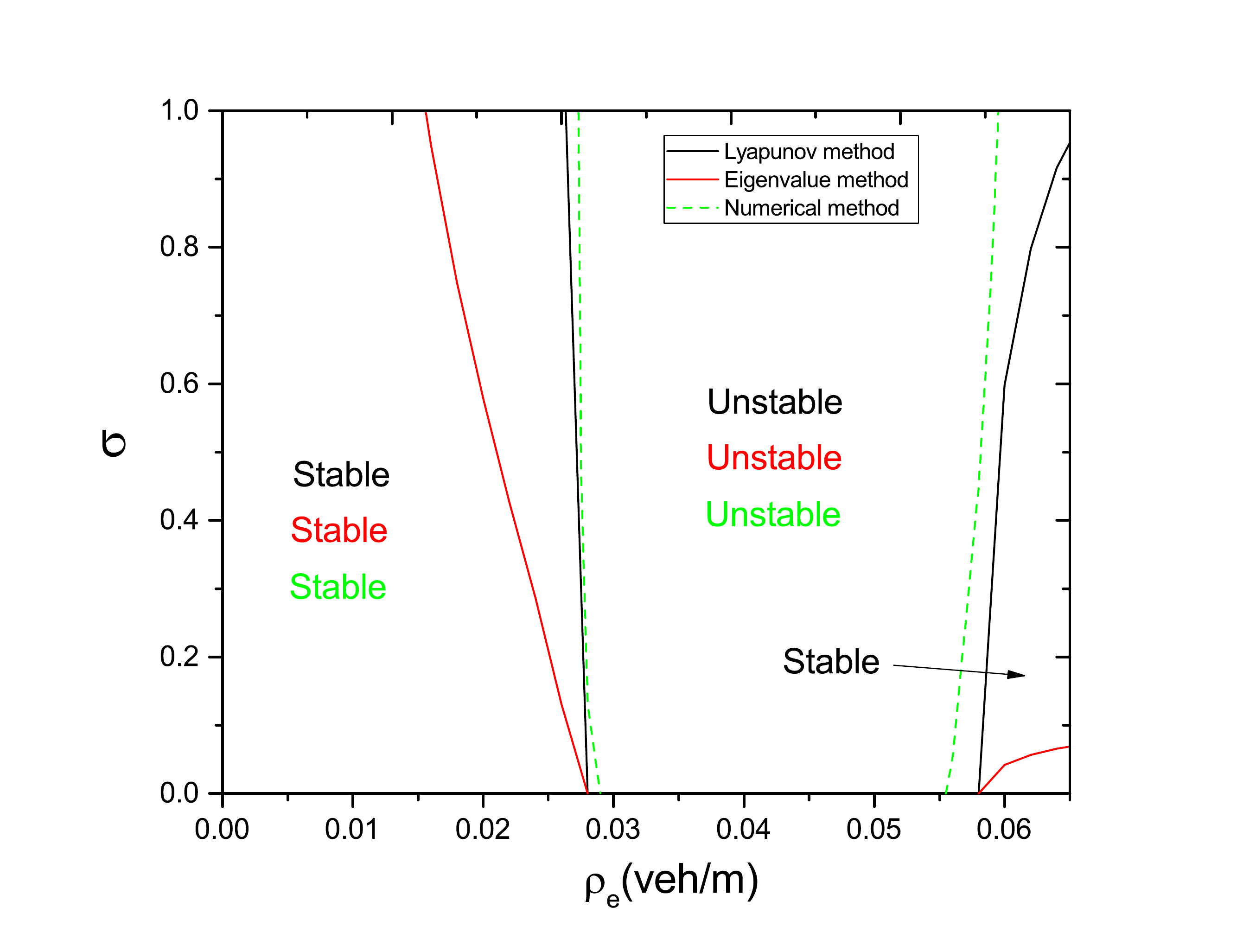}}
\caption{ Stability phase diagrams of the stochastic Aw-Rascle model in the $(\sigma,\rho_{e})$ plane using two stochastic terms, (a) the stochastic term in equation, the unit of $\sigma^2$ is $\mathrm{m/s^{2}}$ (33) (b) the stochastic term in equation (34), the unit of $\sigma^2$ is $\mathrm{m^2/s.veh^{2}}$}.  
\label{fig1}
\end{figure}

Next, we plot the stability phase diagram of the stochastic Aw-Rascle model in the $(\sigma,\rho_{e})$ plane, and compare the theoretical results with numerical simulation for both stochastic terms in equations (33) and (34) (See Figure~\ref{fig1}). For simulation purpose, we choose a traffic pressure having the form $P(\rho)=\alpha \sqrt{\rho}$, where $\alpha=160 SI$ and $\tau=25\ \mathrm{s}$. 

Figure~\ref{fig1}(a) displays the stability diagram for the velocity dependent stochastic term, i.e. equation (33). From the figure, one can clearly see that as we increase the dissipation term $\sigma$, the stable region begins to decrease for high density values. Considering the effect of non-linearity and the numerical integration method, Figure~\ref{fig1}(a) shows that the numerical stability boundary is in an acceptable agreement with the theoretical one. 

Figure~\ref{fig1}(b) displays the stability diagram for the density and velocity dependent stochastic term, i.e. equation (34). Using the LMI solver in Matlab, we solve the LMI in equation (31) by numerically searching a positive defined matrix P for each point in the parameter space. An example to solve the LMI for a point in the parameter space is presented in Appendix C. The stability diagram shows that the stochasticity has an overall less destabilizing effect than in Figure~\ref{fig1}(a), especially for low density values. The numerical simulation is overall in good agreement with our theoretical analysis.

\begin{figure}[H]
\centering
\subfloat[]{\includegraphics[width=.5\textwidth]{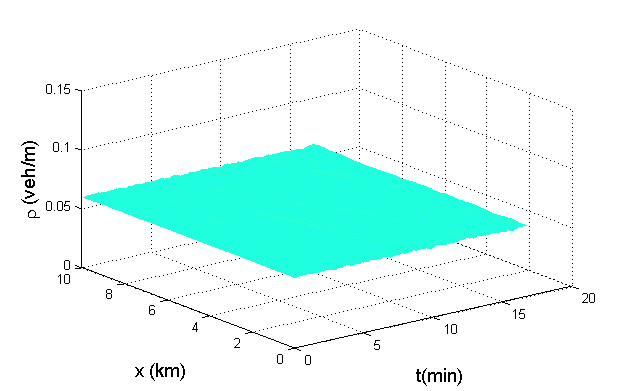}}
\subfloat[]{\includegraphics[width=.5\textwidth]{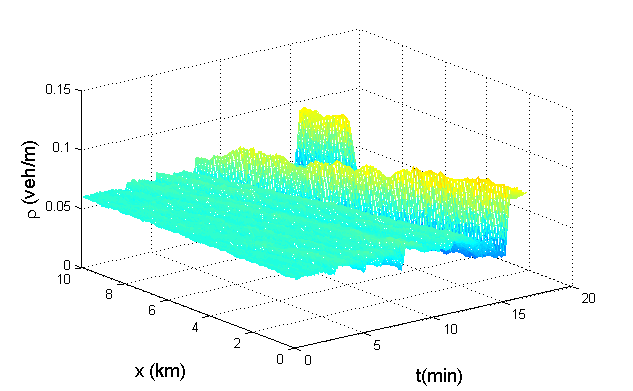}}\\
\caption{Traffic flow evolution of a density $\rho_{e} = 0.06 \ \mathrm{veh/m}$ and two values of $\sigma$, (a) $\sigma^{2}=0.04 \ \mathrm{m/s^{2}}$ (b) $\sigma^{2}=0.64 \ \mathrm{m/s^{2}}$.}
\label{fig2}
\end{figure}

To provide a visual representation of the impact of stochasticity on traffic flow evolution, we plot in Figure~\ref{fig2} the density profiles of $\rho=0.06  \ \mathrm{veh/m}$ and two values of the parameter $\sigma$. Figure~\ref{fig2}(a) shows that traffic is stable under a relatively low value of $\sigma$, and only fluctuations around the equilibrium velocity can be observed. For the same density, Figure~\ref{fig2}(b) shows how stochasticity destabilizes traffic flow when a threshold of $\sigma$ is exceeded (See Figure~\ref{fig1}). Indeed, one can distinguish the emergence of traffic oscillations that are triggered by the presence of stochasticity.

Finally, we compare the stability condition derived in equation (36) with the one derived by \cite{ngodu} which is given by (after replacing in equation (24) in the paper of \cite{ngodu}): 

\begin{equation}
  {(v'_{e} + c_{1})}^{2} \leq c_{2} P'(\rho_{e}) |v'_{e}|, 
\end{equation}

\noindent where $c_{1}=0$ and $c_{2}=1+ \frac{\eta^{2} \tau}{2}$ in the velocity dependent stochastic term. In this case, $\eta=\frac{\sigma}{2\sqrt{ve}}$. For the density and velocity dependent stochastic term, $c_{1}=-\frac{\mu^{2}\tau}{2}$ and $c_{2}=1+ \frac{\eta^{2} \tau}{2}$. In this case, $\mu=\sigma(v_{max}-v_{e})$ and $\eta=-\sigma \rho_{e}$.


From Figure~\ref{fig1}(a), one can see that the stability condition in equation (37) predicts that stochasticity stabilizes the traffic system, which neither conforms with simulations nor with empirical findings. Moreover, the stability condition deviates significantly from numerical simulation. Hence, the stability condition in equation (37) does not agree both qualitatively and quantitatively with numerical simulation for the velocity-dependent stochastic term.  

From Figure~\ref{fig1}(b), the stability condition in equation (37) shows that stochasticity destabilizes traffic performance. Hence, equation (37) is qualitatively in agreement with simulation. However, unlike the result of our analysis, equation (37) underestimates the extension of the stable region. Hence, the Lyapunov method is in better qualitative and quantitative agreement with numerical simulation than the eigenvalue method.


\subsection{The stochastic speed gradient model}

The velocity equation of the speed gradient model \citep{jiang5} has the following expression:  

\begin{equation}
\frac{\partial v}{\partial t} + v \frac{\partial v}{\partial x} = \frac{v_{e}({\rho})-v}{\tau} + c_{0} \frac{\partial v}{\partial x} + f_{2},
\end{equation}

\noindent where $c_{0}$ is the propagation speed of small disturbances. 

\underline{Stability condition}

The stability condition of the stochastic speed gradient model can be obtained by making the following replacement in equation (14): $f_{1vx}=c_{0}$, while the other terms are similar to the stochastic Aw-Rascle model: 

\begin{equation}
c_{0} (2+\tau \eta^2)+2\rho_{e} v'_{e} \geq 0. 
\end{equation}

In the deterministic case ($\eta=0$), the above stability condition becomes $c_{0}+\rho_{e} v'_{e} \geq 0 $. 
For a density and velocity dependent stochastic term, i.e. equation (34), we follow the same methodology explained in the previous section by solving the LMI in equation (31).

\underline{Numerical simulation}

\begin{figure}[H]
\centering
\subfloat[]{\includegraphics[width=.7\textwidth]{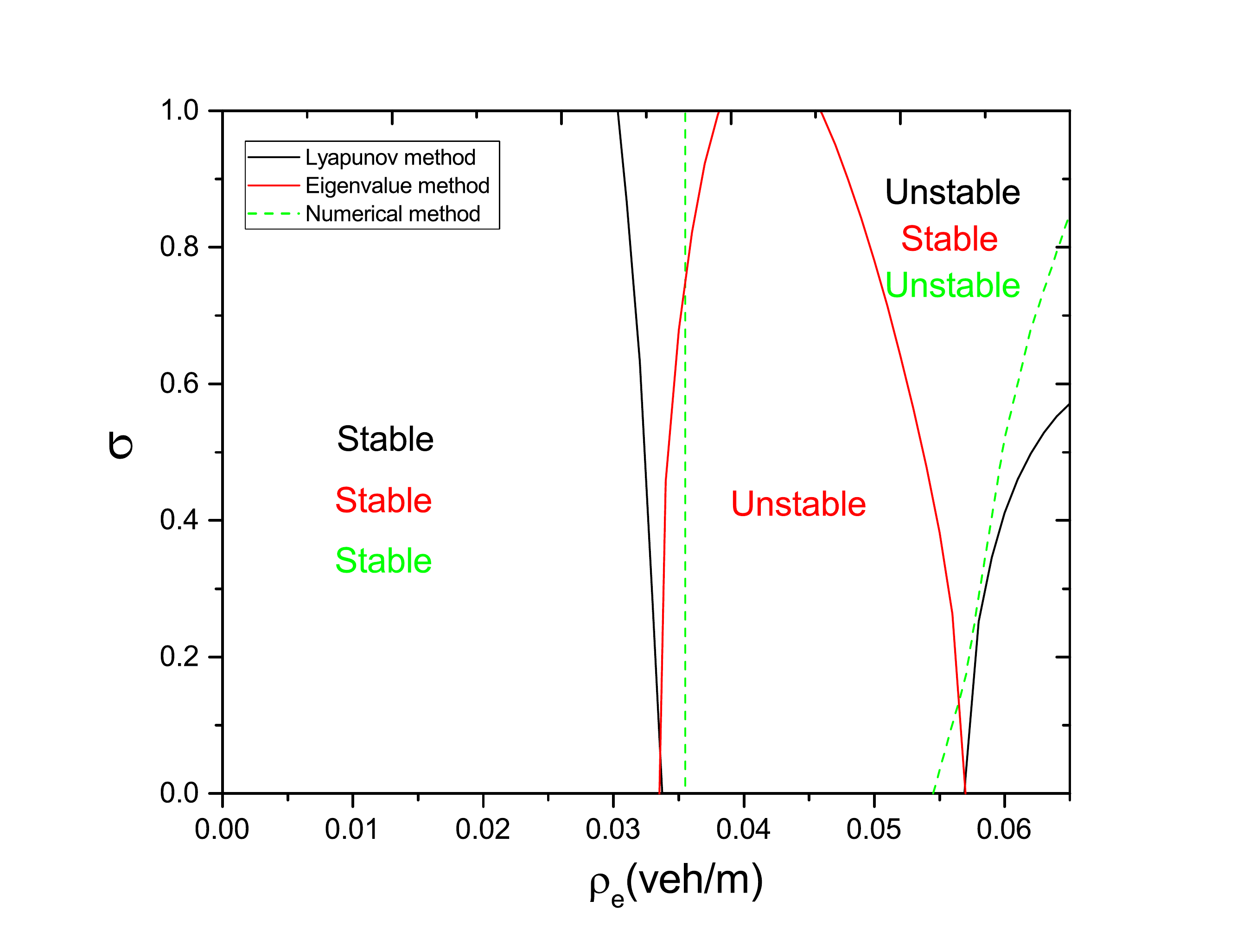}}\\
\subfloat[]{\includegraphics[width=.7\textwidth]{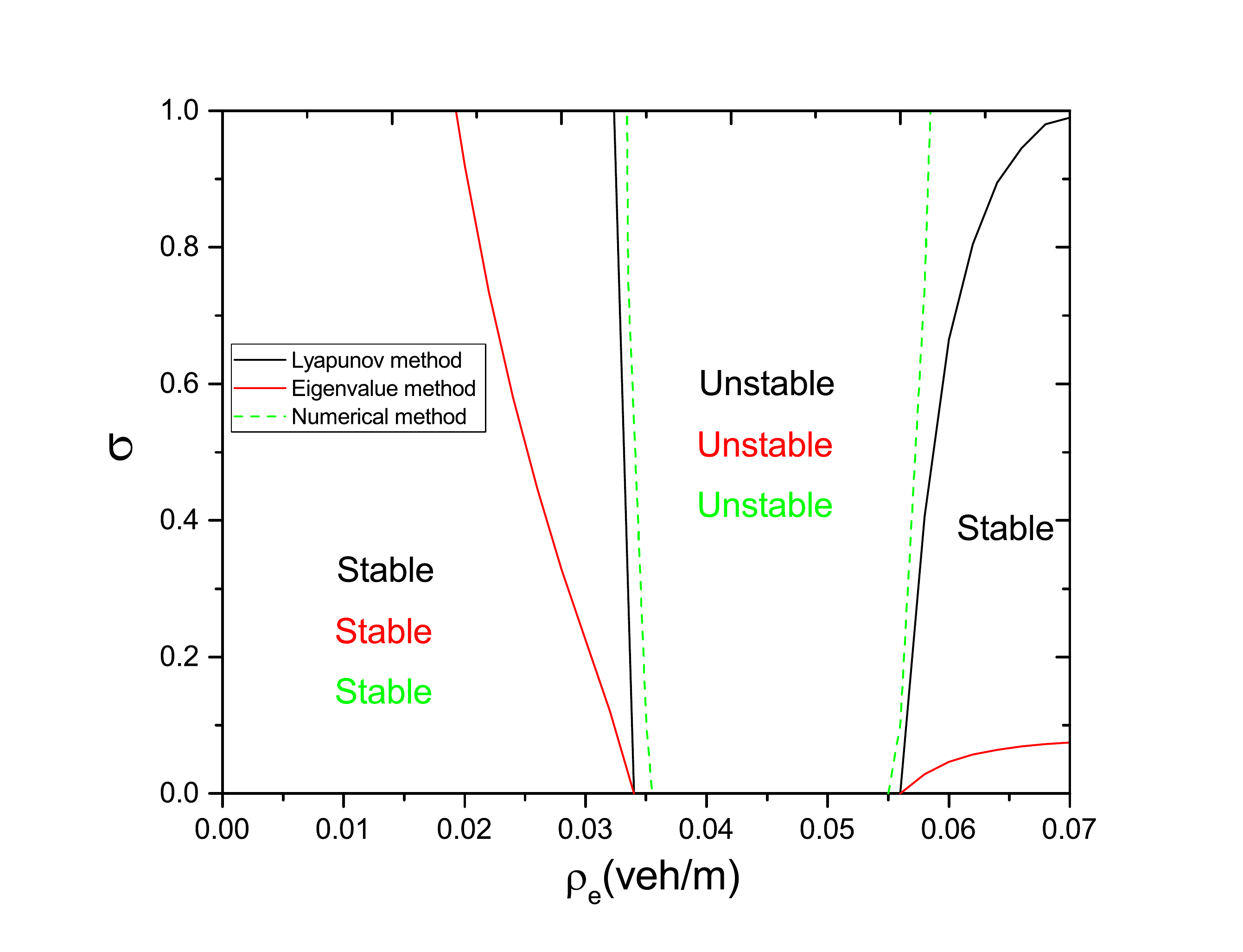}}
\caption{Stability phase diagrams of the stochastic speed gradient model in the $(\sigma,\rho_{e})$ plane using two stochastic terms, (a) the stochastic term in equation (33), the unit of $\sigma^2$ is $\mathrm{m/s^{2}}$. (b) the stochastic term in equation (34), the unit of $\sigma^2$ is $\mathrm{m^2/s.veh^{2}}$.}
\label{fig3}
\end{figure}

Figure~\ref{fig3} displays the phase diagram of the stochastic speed gradient model in the $(\sigma,\rho_{e})$ plane for $c_{0}=20 \ \mathrm{m/s}$ and $\tau=25\ \mathrm{s}$. Figure~\ref{fig3}(a) and Figure~\ref{fig3}(b) correspond to the stochastic term in equation (33) and (34), respectively.  Similar to the Aw-Rascle model, Figure~\ref{fig3}(a) shows that the stochasticity has a significant destabilizing effect for high density values, while the effect of stochasticity is mild for low density values.  Figure~\ref{fig3}(b) shows that the stochasticity has a small destabilizing effect compared with Figure~\ref{fig3}(a). The difference in the impact of stochasticity using equations (33) or (34) is more pronounced when the density is high. 

The numerical stability boundary is in good agreement with our theoretical analysis. The difference enlarges for high density values, which is probably due to the high level of noise compared with the equilibrium velocity and the effect of non-linearity. Note that when the density is higher, the value of $\sigma$ in Figure~\ref{fig3} becomes very high compared with the equilibrium velocity, which is not realistic.

\begin{figure}[H]
\centering
\subfloat[]{\includegraphics[width=.5\textwidth]{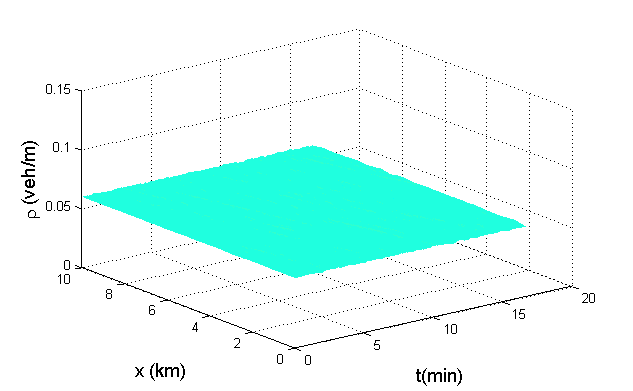}}
\subfloat[]{\includegraphics[width=.5\textwidth]{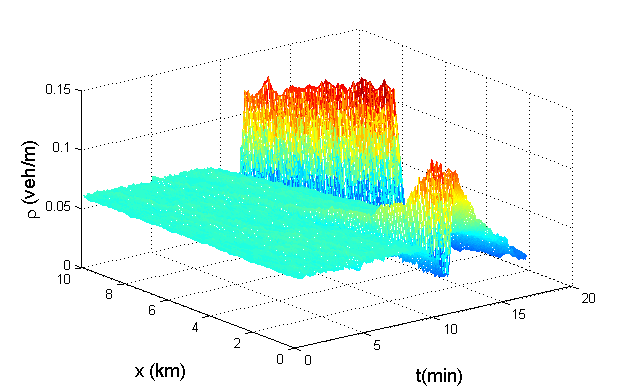}}\\
\caption{Traffic flow evolution of a density $\rho_{e} = 0.06 \ \mathrm{veh/m}$ and two values of $\sigma$, (a) $\sigma^{2}=0.04 \ \mathrm{m/s^{2}}$ (b) $\sigma^{2}=0.64 \ \mathrm{m/s^{2}}$.}
\label{fig4}
\end{figure}

Next, we plot in Figure~\ref{fig4} the density profile of $\rho=0.06  \ \mathrm{veh/m}$ and two values of the parameter $\sigma$. From the Figures, one can see that instability is triggered by stochasticity alone when a given threshold of $\sigma$ is exceeded for high density values. 

The stability condition derived by \cite{ngodu} is given by:

\begin{equation}
  \rho_{e}{(v'_{e} + c_{1})}^{2} \leq c_{2} c_{0} |v'_{e}|, 
\end{equation}

\noindent where the coefficients $c_{1}$ and $c_{2}$ are the same as the stochastic Aw-Rascle model. 
As observed in the stochastic Aw-Rascle model, the stability condition in equation (40) neither conforms with the results of numerical simulations nor the Proposition 1 in \cite{ngodu} for the velocity dependent stochastic term. However, for the density and velocity dependent stochastic term, equation (40) agrees with simulation only qualitatively. 

\subsection{The stochastic Zhang model}

The velocity equation of the stochastic Zhang model has the following expression:  

\begin{equation}
\frac{\partial v}{\partial t} + v \frac{\partial v}{\partial x} = \frac{v_{e}({\rho})-v}{\tau} - \rho v'_{e} \frac{\partial v}{\partial x} + f_{2},
\end{equation}

\underline{Stability condition}

The parameters of the Zhang model lead to a singular Matrix P in equation (25) since the resulting eigenvalue in equation (27) is zero. Hence, the matrix P in equation (25) is not positive defined. In this case, traffic stability cannot be mathematically demonstrated by the present approach.

\underline{Numerical simulation}

\begin{figure}[H]
\centering
\subfloat[]{\includegraphics[width=.5\textwidth]{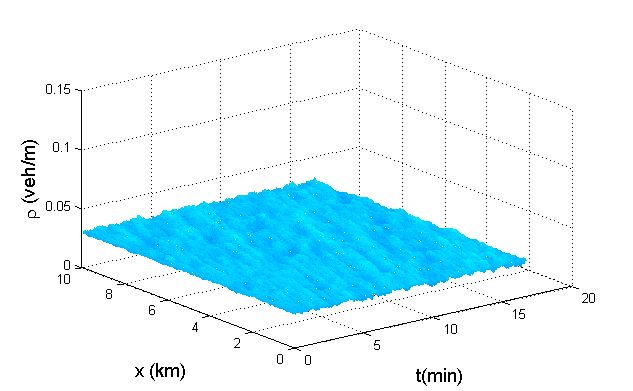}}
\subfloat[]{\includegraphics[width=.5\textwidth]{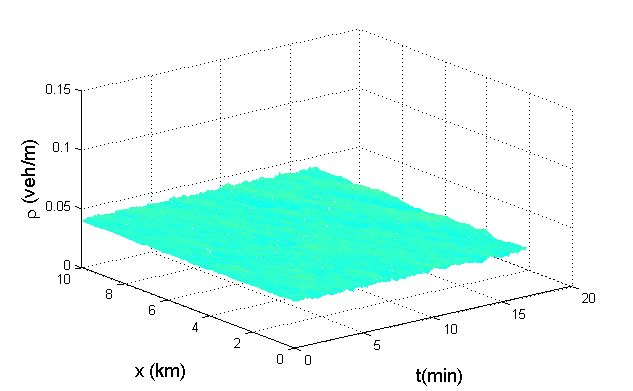}}\\
\caption{Traffic flow evolution of two density values with $\sigma^{2}=0.64 \ \mathrm{m/s^{2}}$, (a) $\rho=0.03 \ \mathrm{veh/m}$ (b) $\rho=0.04 \ \mathrm{veh/m}$.}
\label{fig5}
\end{figure}

The numerical simulation of the deterministic model shows that traffic is stable. Moreover, no destabilizing effect of stochasticity has been reported for both stochastic terms in equations (33) and (34). Figure~\ref{fig5}(a,b) show the density profiles corresponding to two density values and a relatively high value of $\sigma$. These densities correspond to the most unstable region in the above stochastic models, where traffic is unstable even in the deterministic case. One can see that stochasticity does not have a qualitative effect on traffic stability.

For the velocity dependent stochastic term, the stability condition derived by \cite{ngodu} yields to: 

\begin{equation}
\eta \geq 0,
\end{equation}
 
\noindent which qualitatively and quantitatively agrees with numerical simulation. However, for the density and velocity dependent stochastic term, the stability condition derived by \cite{ngodu} yields to:

\begin{equation}
  {(v'_{e} + c_{1})}^{2} \leq c_{2} {|v'_{e}|}^{2}, 
\end{equation}

\noindent which is always not satisfied for $\sigma>0$. Hence, equation (43) predicts that traffic is unstable which does not agree with simulation qualitatively and quantitatively. 

\subsection{The stochastic gas kinetic traffic model}

\cite{prigo} have been the first to propose a gas kinetic traffic (GKT) model which has been further studied and improved \citep{helb1,helb2}. In particular, \cite{trb} proposed a non-local gas kinetic model where an interaction distance has been taken into account. In this case, the velocity equation of the stochastic GKT model reads \citep{trbb}: 

\begin{equation}
\frac{\partial v}{\partial t}+v\frac{\partial v}{\partial x}=-\frac{1}{\rho} \frac{\partial (\rho \theta)}{\partial x} + \frac{v^{*}_{e}-v}{ \tau} + f_{2},
\end{equation}


\noindent where

\begin{equation}
v^{*}_{e} = v_{0}[1-\frac{A(\rho)}{A(\rho_{m})}{(\frac{\rho_{a}vT}{1- \frac{\rho_{a}}{\rho_{m}}})}^{2}B(\delta_{v})],
\end{equation}

\begin{equation}
B(\delta_{v}) = 2[\delta_{v} \frac{e^{-\delta_{v}^2}}{\sqrt{2\pi}}+(1-\delta_{v}^2)\int_{-\infty}^{\delta_{v}}{\frac{e^{-y^2/2}}{\sqrt{2\pi}}}], 
\end{equation}

\begin{equation}
\delta_{v}=\frac{(v-v_{a})}{\sqrt{\theta}},
\end{equation}

\noindent The velocity variance $\theta$ is assumed to be proportional to $v^{2}$ \citep{trb}: 

\begin{equation}
\theta = A(\rho) v^{2}, 
\end{equation}

\noindent where $A(\rho)$ is the variance prefactor, T is the safe time headway, $\rho_{a}$ and $v_{a}$ are the density and the velocity at $x_{a} = x + d$, $d=Tv$ is the interaction distance and $v_{0}$ is the maximum velocity. 

For the variance prefactor $A$, we adopt the following formula \citep{trb,trbb}:  

\begin{equation}
A{\rho} = \alpha_{f} + \frac{\alpha_{c}-\alpha_{f}}{2}[1+\tanh{(\frac{\rho-\rho_{c}}{\delta \rho})}].
\end{equation}

\underline{Stability condition}

From the model defined above, the derivatives intervening in the stability condition in equation (14) are given by the following:


\begin{equation}
f_{1\rho x} = -\frac{1}{\rho_{e}}(\theta+\rho_{e}A'(\rho)v^{2}_{e}), 
\end{equation}

\begin{equation}
f_{1\rho a} = \frac{-2(v_{0}-v_{e})\rho_{m}}{\tau \rho_{e}(\rho_{m}-\rho_{e})}, 
\end{equation}

\begin{equation}
f_{1va} = \frac{2(v_{0}-v_{e})}{\tau \sqrt{\pi \theta}}, 
\end{equation}

\begin{equation}
f_{1\rho} = \frac{(v_{e}-v_{0})A'(\rho)}{\tau A(\rho)},
\end{equation}

\begin{equation}
f_{1v} = \frac{-f_{\rho}-f_{\rho a}-f_{va} f_{\rho} \tau}{f_{\rho} \tau},
\end{equation}


\underline{Numerical simulation}

\begin{figure}[H]
\centering
\subfloat[]{\includegraphics[width=.7\textwidth]{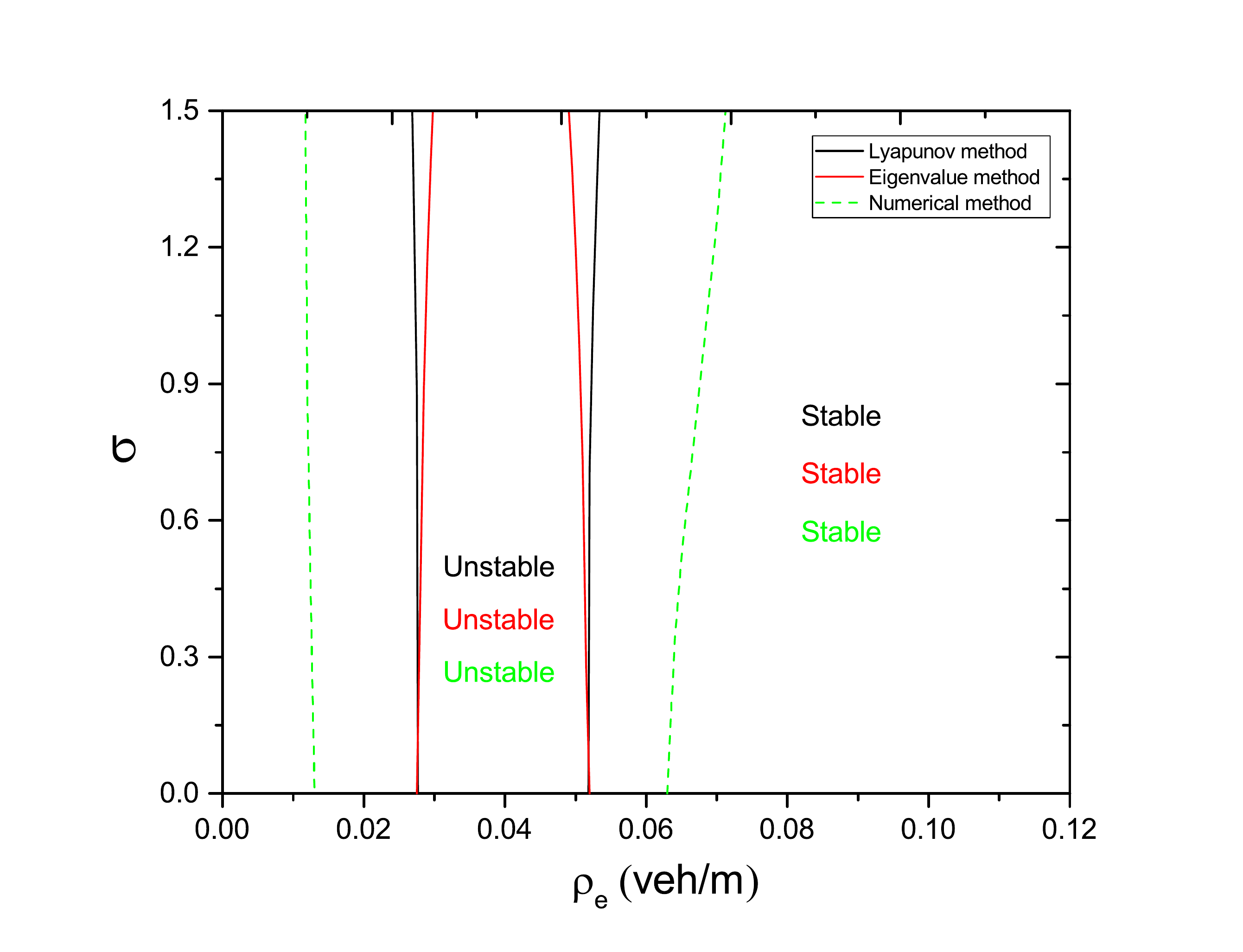}}\\
\subfloat[]{\includegraphics[width=.7\textwidth]{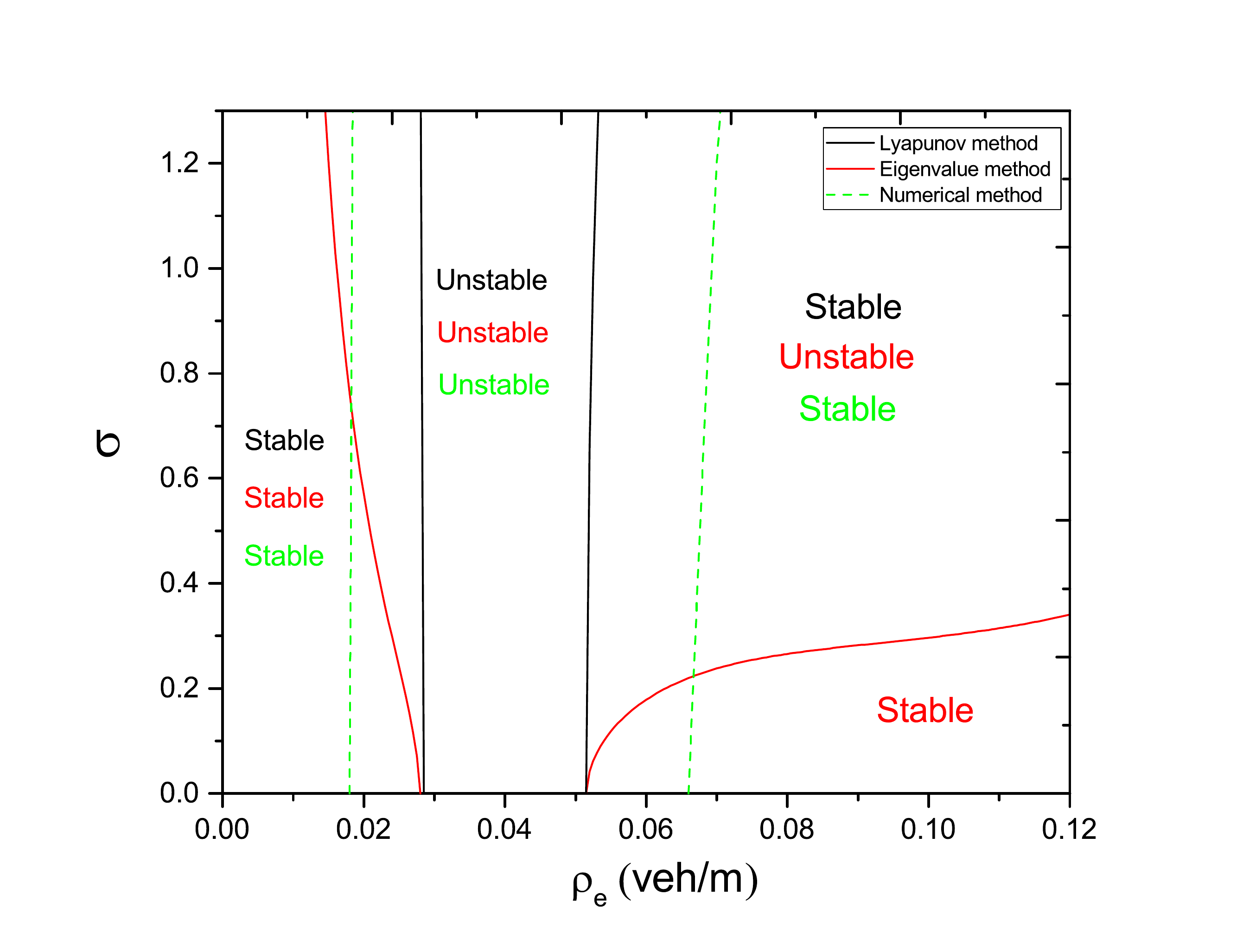}}
\caption{ Stability phase diagrams of the stochastic gas kinetic traffic model in the $(\sigma,\rho_{e})$ plane using two stochastic terms, (a) the stochastic term in equation (33), the unit of $\sigma^2$ is $\mathrm{m/s^{2}}$. (b) the stochastic term in equation (34), the unit of $\sigma^2$ is $\mathrm{m^{2}/s.veh^{2}}$.}
\label{fig6}
\end{figure}

Figure~\ref{fig6}(a) shows the stability diagram corresponding to the stochastic GKT model using the velocity dependent stochastic term in equation (33). In this context, the following parameters were used, $T=1.2$ s, $\rho_{m}=0.16 \ \mathrm{veh/m}$, $v_{0}=33 \ \mathrm{m/s}$, $\tau=25$s, $\alpha_{f}=0.01$, $\alpha_{c}=0.04$, $\rho_{c}=0.03 \ \mathrm{veh/m}$ and $\delta \rho =0.15 \rho_{m} \ \mathrm{veh/m}$ \citep{ngodu}. From Figure~\ref{fig6}(a), one can see that traffic is stable for both low and high densities while it is unstable for a given range of intermediate densities. The stability condition in equation (14) predicts that traffic gets unstable with the increase of $\sigma$. The unstable region increases to include densities from the left and the right. However, compared with the stochastic Aw-Rascle model and the stochastic speed gradient model, the stochasticity has a small influence on traffic stability for high density values.

Figure~\ref{fig6}(b) displays the stability diagram corresponding to the stochastic GKT model using the density and velocity dependent stochastic term in equation (34). One can see that the stability diagram predicts that stochasticity impacts only a very limited density range. 

Next, we plot the numerical boundary corresponding to simulation results for different values of $\sigma$. The numerical simulation shows that stochasticity destabilizes traffic as we increase the parameter $\sigma$, which is in agreement with analytical results. The quantitative difference between theory and simulation is more pronounced than the previously studied stochastic models, which is probably related to the more complex acceleration function of the GKT model. 

\begin{figure}[H]
\centering
\subfloat[]{\includegraphics[width=.5\textwidth]{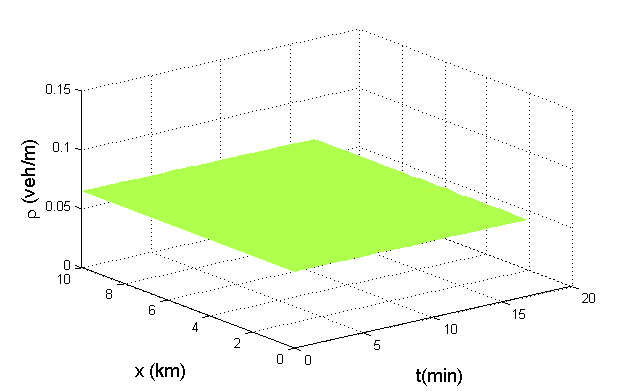}}
\subfloat[]{\includegraphics[width=.5\textwidth]{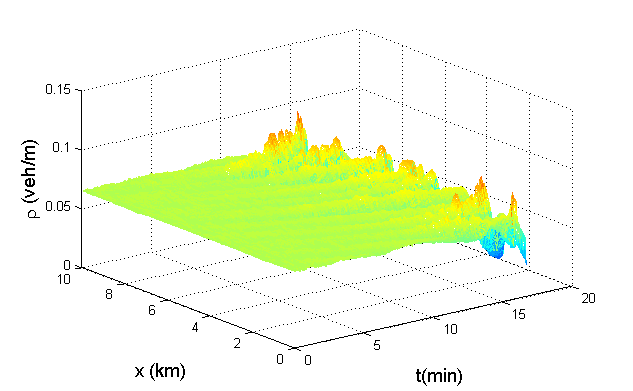}}\\
\caption{Traffic flow evolution of a density $\rho_{e} = 0.065 \ \mathrm{veh/m}$ and different values of $\sigma$, (a) $\sigma^{2}=0.04 \ \mathrm{m/s^{2}}$ (b) $\sigma^{2}=0.64 \ \mathrm{m/s^{2}}$.}
\label{fig7}
\end{figure}

To provide a visual representation of the instability induced by stochasticity, we plot in Figure~\ref{fig7}, the density profiles corresponding to a fixed density $\rho_{e}= 0.065 \ \mathrm{veh/m}$ and two different values of $\sigma$. Figure~\ref{fig7}(a) shows that a relatively small value of $\sigma$ causes fluctuations around the equilibrium, but does not destabilize traffic flow. Figure~\ref{fig7}(b) shows how traffic instability can be induced by stochasticity for a relatively high value of $\sigma$.

   
Finally, we compare the stability condition in equation (14) with the stability condition derived by \cite{ngodu} which is given by: 

\begin{equation}
(\rho_{e} (f_{1\rho} \tau+c_{1})^2 - c_{2}(\theta+\rho_{e}A'(\rho)v^{2}_{e} - \frac{\rho_{e}d}{\tau}(f_{1\rho} \tau^{2} f_{1va} + f_{1\rho a}\tau)) \leq 0
\end{equation}





\noindent where $c_{1}=0$ and $c_{2}=1+ \frac{\eta^{2}\tau v_{e}}{2(2v_{0}-v_{e})}$ in the velocity dependent stochastic term. In this case, $\eta=\frac{\sigma}{2\sqrt{ve}}$. For the density and velocity dependent stochastic term, $c_{1}=-\frac{\mu^{2} \tau v_{e}}{2(2v_{0}-v_{e})}$ and $c_{2}=1+ \frac{\eta^{2} \tau v_{e}}{2(2v_{0}-v_{e})}$. In this case, $\mu=\sigma(v_{0}-v_{e})$ and $\eta=-\sigma \rho_{e}$.

As shown in Figure~\ref{fig6}(a), the stability condition in equation (55) predicts that stochasticity improves traffic performance as we increase the parameter $\sigma$. The stability condition in equation (55) neither conforms with numerical simulations nor with empirical findings. 

\begin{figure}[H]
\centering
\subfloat[]{\includegraphics[width=.5\textwidth]{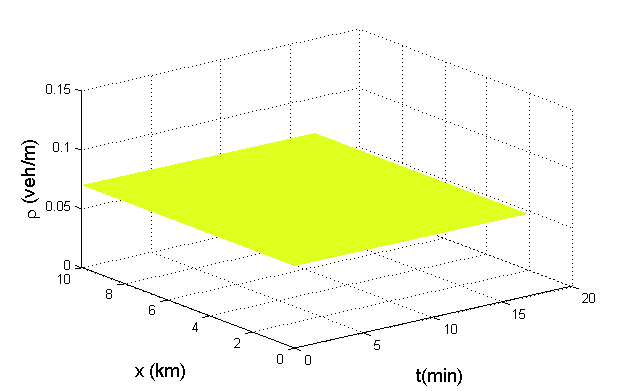}}
\subfloat[]{\includegraphics[width=.5\textwidth]{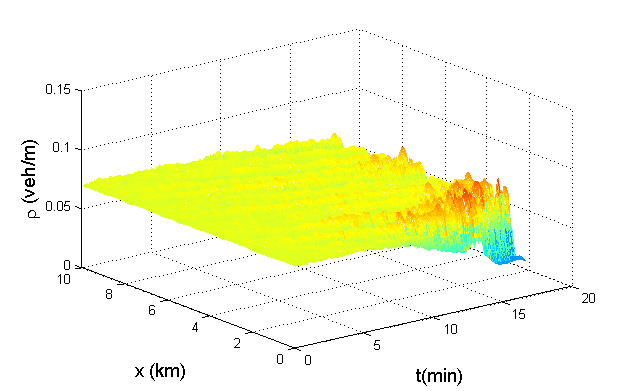}}\\
\caption{Traffic flow evolution of a density $\rho_{e} = 0.07 \ \mathrm{veh/m}$ and different values of $\sigma$, (a) $\sigma^{2}=0 \ \mathrm{m^{2}/s.veh^{2}}$ (b) $\sigma^{2}=1.7 \ \mathrm{m^{2}/s.veh^{2}}$.}
\label{fig8}
\end{figure}

\begin{figure}[H]
\centering
\includegraphics[width=.6\textwidth]{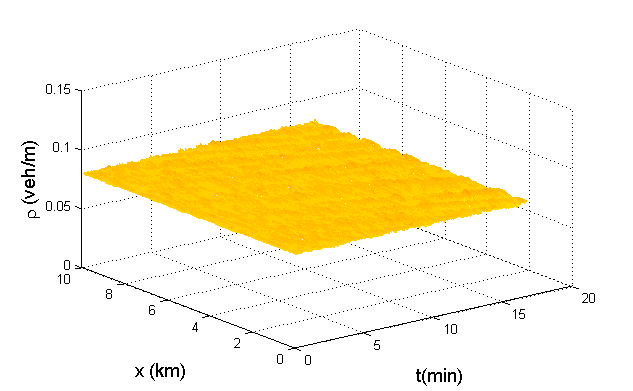}
\caption{Traffic flow evolution of a density $\rho_{e} = 0.08 \ \mathrm{veh/m}$ and $\sigma^{2}=0.64 \ \mathrm{m^{2}/s.veh^{2}}$.}
\label{fig9}
\end{figure}

For the density and velocity-dependent stochastic term in equation (34), one can see that the stability condition derived by \cite{ngodu} can predict the destabilizing effect of stochasticity, see Figure~\ref{fig6}(b). However, equation (55) largely underestimates the extension of the stable region. In this case, stochasticity has only a mild effect on traffic stability. To provide a visual representation, we plot in Figure~\ref{fig8} with $\Delta t =0.02$ s, the density profiles corresponding to a density $\rho_{e} = 0.07 \ \mathrm{veh/m}$ to show the noise-induced instability. As shown in \cite{ngodu}, stochasticity indeed destabilizes traffic flow. However, as we move to a higher density, one can see that stochasticity does not induce instability for reasonable values of $\sigma$, see for instance Figure~\ref{fig9} for a density $\rho_{e} = 0.08 \ \mathrm{veh/m}$.

\section{Conclusion}

It is well known that the presence of noise can either stabilize or destabilize physical systems. In traffic flow, recent empirical investigations have demonstrated that stochastic factors play an important role in destabilizing traffic flow.  In this study, we have conducted a stability analysis of a general stochastic continuum second-order macroscopic model and demonstrated that the stochastic factors have a qualitative influence on traffic stability.
 
The stochastic macroscopic models respond differently to the presence of stochasticity. On the other hand, except the Zhang model where stochasticity seems to not have an influence on traffic stability, the other traffic models with a speed gradient in the velocity equation have shown a significant destabilizing effect of stochasticity when the traffic density is high, while on the other hand, the stochasticity has shown a mild influence on traffic stability for the GKT model.  

From the methodological perspective, we have also shown that the direct Lyapunov method represents a simple and efficient tool to study the stability of stochastic macroscopic traffic models. The Lyapunov method can be extended in future works to derive stability conditions for more improved traffic models.

Nevertheless, further investigations should be performed for more accurate modeling of human stochastic aspects and more complex traffic situations. Finally, this work may offer a new perspective on traffic management and control by considering the stochastic nature of human drivers from a macroscopic level.




\hspace{10pt}

\begin{flushleft}
\normalsize{\bfseries{\large{Acknowledgments}}}
\end{flushleft}

\hspace{10pt}
This work was supported by the National Key R \& D Program of China (No. 2018YFB1600900), the National Natural Science Foundation of China (No. 71621001, 71971015 and 71931002).

\bibliography{citation}

\begin{appendices}
\counterwithin{figure}{section}
\counterwithin{table}{section}
\counterwithin{equation}{section}


\section{Existence and uniqueness of the solution}

\begin{myDef1}
     
Consider the following stochastic differential equation:

\begin{equation}
dX=f(x,t)dt+h(x,t)dW.
\end{equation}

\noindent (i) The solution of (A.1) exists if the following condition is satisfied (linear growth) \citep{zhang3}: 

\begin{equation}
\lVert f(x,t) \rVert +\lVert h(x,t) \rVert \leq \lambda_{1}(1+\lVert x \rVert).
\end{equation}

\noindent This condition guarantees that the solution does not blow up from a given time step, i.e. the solution does not have finite time escape.

\noindent (ii) The solution of (A.1.) is unique if the following condition is satisfied (Lipschitz continuity) \citep{zhang3}:  

\begin{equation}
\lVert f(x,t)-f(y,t) \rVert + \lVert h(x,t)-h(y,t) \rVert \leq \lambda_{2} \lVert x-y \rVert.
\end{equation}

\noindent $\lambda_{1}$ and $\lambda_{2}$ are positive constants and $\lVert.\rVert$ denotes the Euclidean norm. 

\end{myDef1}

We apply the above definition to the stochastic continuum macroscopic model in equations (1) and (2) by considering $f(x,t)=(s,f_{1})$, $s=-\frac{\partial\rho v}{\partial x}$, $h(x,t)=(0,f_{2})$ and $x=(\rho,v)$.

\noindent (i) Existence:

The deterministic model is hyperbolic which ensures the existence of solution given an initial condition; hence, the Euclidian norm $\lVert f(x,t) \rVert$ is finite, namely, $\lVert f(x,t) \rVert \leq C$. Moreover, assume $\lVert h(x,t) \rVert \le \sigma \lVert x \rVert$, Hence:

\begin{equation}
\lVert f(x,t) \rVert +\lVert h(x,t) \rVert \leq C+\sigma \lVert x \rVert =C(1+\frac{\sigma}{C} \lVert x \rVert).
\end{equation}

\noindent Consequently, the following inequality:

\begin{equation}
C(1+\frac{\sigma}{C}\lVert x\rVert)\leq C(1+\lVert x \rVert),
\end{equation}

\noindent is guaranteed if $\frac{\sigma}{C}\le1$.
The dissipation parameter $\sigma$ should be smaller than the upper bound of the Euclidian norm of the function $f$ to guarantee existence.

\noindent (ii) Uniqueness

The functions $f$ and $h$ are both continuous and differentiable; hence, after applying the intermediate value theorem, we get: 

\begin{equation}
\lVert f(x,t)-f(y,t) \rVert \leq m_{1} \lVert x-y \rVert,
\end{equation}

\begin{equation}
\lVert h(x,t)-h(y,t) \rVert \leq m_{2} \lVert x-y \rVert,
\end{equation}

\noindent The summation of the above inequalities yield:

\begin{equation}
\lVert f(x,t)-f(y,t) \rVert+ \lVert h(x,t)-h(y,t) \rVert \leq \lambda_{2} \lVert x-y \rVert,
\end{equation}

\noindent where $\lambda_{2}=m_{1}+m_{2}$. Consequently, the solution is unique.



\section{First order upwind scheme}

In numerical simulations, we have applied the numerical integration method firstly proposed by \cite{jiang5} for the deterministic speed gradient model. For the stochastic component, we have adopted the Euler-Maruyama integration scheme \citep{mur}.

\noindent For density equation update, the following discretization form has been used:  

\begin{equation}
\tilde{\rho}_{i}^{j+1}= \rho_{i}^{j} - \frac{dt}{dx} (v_{i+1}^{j}-v_{i}^{j}) \rho_{i}^{j}- \frac{dt}{dx} v_{i}^{j} (\rho_{i}^{j}-\rho_{i-1}^{j}).
\end{equation}

It has been demonstrated that the equation (B.1) ensures the suitable physical direction of the information propagation \citep{jiang5}.

\noindent For the velocity equation, the following integration scheme is adopted:

\noindent if $v_{i}^{j}<c_{0}$: 

\begin{equation}
\begin{multlined}
\tilde{v}_{i}^{j+1}= v_{i}^{j} - \frac{dt}{dx} (v_{i}^{j}-c_{0})(v_{i+1}^{j} - v_{i}^{j}) - dt\frac{(v_{i}^{j}-v_{e})}{\tau} - \sigma \sqrt{v_{i}^{j}}dB_{i}^{j}(t). 
\end{multlined} 
\end{equation}

\noindent if $v_{i}^{j}\geq c_{0}$: 

\begin{equation}
\begin{multlined}
\tilde{v}_{i}^{j+1}= v_{i}^{j} - \frac{dt}{dx}(v_{i}^{j}-c_{0})(v_{i}^{j} - v_{i-1}^{j}) - dt\frac{(v_{i}^{j}-v_{e})}{\tau} + \sigma \sqrt{v_{i}^{j}}dB_{i}^{j}(t). 
\end{multlined} 
\end{equation}


\noindent where $dB_{i}^{j}(t)$ is a normal distribution of mean 0 and standard deviation $\sqrt{dt}$.

\section{Example of solving the LMI}

We present an example using the stochastic Aw-Rascle model in equation (35) with a density and velocity dependent stochastic term, see equation (34). 

Consider a traffic density of $\rho_{e} = 0.07 \ \mathrm{veh/m}$, $\sigma^{2} = 0.25 \ \mathrm{m/s^{3}}$ and $k = 0.01 \ \mathrm{veh^{-1}}$. Note that in the stability analysis, we have performed a long-wave length approximation to estimate the matrix $A_{s}$, hence $k \rightarrow 0$. Accordingly, the stability diagram will not be influenced when choosing small values of $k \rightarrow 0$, e.g. $k \leq 0.25$ \citep{trbb}. After using the default parameters reported in section 4.1 for the stochastic Aw-Rascle model, we obtain the following expression of the matrix $A_{s}$ in equation (17):

\begin{equation}
A_{s}=  {\begin{bmatrix}
      0     & 0 & -0.0279 & -0.0007 \\
  -6.6524    &  -0.0400 &  0 &  0.1838  \\
	 0.0279   &  0.0007   &   0 & 0  \\
	0   &   -0.1838  & -6.6524 & -0.0400  \\
\end{bmatrix}}.
\end{equation}

\noindent Based on equation (10) and equation (34), the matrix $R_{s}$ in equation (18) will be the following:

\begin{equation}
R_{s}=  {\begin{bmatrix}
      0     & 0 & 0 & 0\\
  13.6069   &  -0.0350 & 0 &  0  \\
	0   &  0  &   0 & 0  \\
	0   &   0 &   13.6069 & -0.0350  \\
\end{bmatrix}}.
\end{equation}

Next, we run the LMI solver algorithm in Matlab software and we get the following matrix:

\begin{equation}
P=  10^{4} \times {\begin{bmatrix}
      6.9929     & 0.0167 & 0 & -0.0283 \\
  0.0167    &  0.0008 & 0.0283 & 0  \\
	0   &  0.0283  &    6.9929 & 0.0167  \\
	-0.0283   &  0 & 0.0167 & 0.0008  \\
\end{bmatrix}}, 
\end{equation}
   
The matrix P in equation (C.3) is a positive defined matrix since its eigenvalues are positive. Hence, traffic is stable for $\rho = 0.07 \ \mathrm{veh/m}$ and $\sigma^{2} = 0.25 \ \mathrm{m/s^{3}}$. In the case where a matrix P satisfying the inequality (31) is not found, the problem is called infeasible and the traffic is not stable.

\end{appendices}

\end{document}